# Do Truly Unidirectional Surface Plasmon-Polaritons Exist?

S. ALI HASSANI GANGARAJ*, FRANCESCO MONTICONE*

*Cornell University, School of Electrical and Computer engineering, Ithaca, New York, 14853, USA*
*Corresponding authors: ali.gangaraj@gmail.com ; francesco.monticone@cornell.edu*

In this work, we revisit the topic of surface waves on nonreciprocal plasmonic structures, and clarify whether strictly unidirectional surface plasmon-polaritons are allowed to exist in this material platform. By investigating different three-dimensional configurations and frequency regimes, we theoretically show that, while conventional surface magneto-plasmons are not strictly unidirectional due to nonlocal effects, consistent with recent predictions made in the literature, another important class of one-way surface plasmon-polaritons, existing at an interface with an opaque isotropic material, robustly preserve their unidirectionality even in the presence of nonlocality, and for arbitrarily-small levels of dissipation. We also investigate the extreme behavior of *terminated* unidirectional wave-guiding structures, for both classes of surface waves, and discuss their counter-intuitive implications.

## 1. INTRODUCTION

Unidirectional surface plasmon-polaritons (SPPs) on nonreciprocal material platforms are currently the subject of significant interest and some controversy. It has been known for several decades that, at the interface between a magnetized plasma (a gyrotropic medium) and an isotropic material, unidirectional surface waves can emerge under certain conditions and in certain frequency ranges (e.g., [1-3]). This topic has seen a recent resurgence of interest in the context of nonreciprocal and topological electromagnetics [4-16]. Particularly intriguing is the possibility of realizing truly unidirectional wave-propagation channels, combined with the high degree of field localization and confinement of plasmonic platforms, which may enable extreme and counter-intuitive optical effects, especially if the one-way channel is abruptly terminated or closed. Notably, an apparent violation of the time-bandwidth limit in a nonreciprocal cavity based on magnetized plasmonic materials has been recently reported in [11], and later clarified in [12,13,14]. The difficulties in correctly interpreting the extreme wave-propagation effects in nonreciprocal plasmonics are also associated with the ubiquity of apparent thermodynamic paradoxes in these systems, some of which were already discussed in the 1960s [2,17].

Interestingly, an important class of nonreciprocal SPP modes has been recently shown to not exhibit strict unidirectionality if more realistic material models that include nonlocal effects (spatial dispersion) are considered [14], challenging widely accepted ideas about nonreciprocal plasmonic systems. However, it is still unclear whether nonlocal effects prevent the unidirectionality of surface plasmon-polaritons in all cases, or whether truly unidirectional (and topologically-protected) SPPs may exist in different regimes of nonreciprocal plasmonics.

In this Article, we address these relevant issues. We show that the effect of nonlocality strongly depends on the considered structure. While in certain frequency regimes and configurations, SPP modes clearly lose their unidirectionality due to nonlocal and other effects, spatial dispersion does not always prevent the existence of unidirectional surface waves, even in the lossless case. As we discuss in the following, this form of unidirectionality does not lead to thermodynamic paradoxes, but it may produce extremely localized and enhanced fields in suitable configurations.

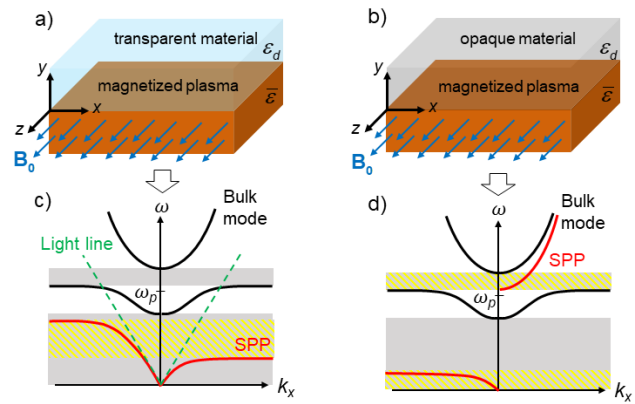

Fig. 1. Nonreciprocal plasmonic platforms. (a,b) Two configurations that support surface plasmon-polaritons (SPPs) between a magnetized plasma and (a) a transparent or (b) an opaque isotropic material. (c,d) Illustration of the typical dispersion diagrams for the configurations in (a) and (b), respectively. The plots indicate the bulk modes (solid black curves), bound SPP modes (solid red), and light line (dashed green), as well as the bulk-mode bandgaps (shaded gray areas), and unidirectional frequency windows (yellow-gray areas) for the case of local materials.

## 2. NONRECIPROCAL PLASMONIC PLATFORMS

Different forms of bulk-wave and surface-wave propagation exist in a three-dimensional nonreciprocal plasmonic structure, depending on the material configuration and frequency. In this Article, we investigate all the scenarios that appear to lead to unidirectional surface-wave propagation, keeping our discussion as general as possible. As illustrated in Fig. 1, the geometry of interest is a stratified structure along the *y*-axis, transversely-invariant in the *xz*-plane, with an interface between a nonreciprocal gyrotropic material in the region $y < 0$, and an isotropic material for $y \geq 0$.

The gyrotropic medium is a plasma or plasmonic material biased by a static magnetic field along the *z*-axis, which breaks reciprocity and time-reversal symmetry [18]. As a result, the permittivity of the magnetized plasma becomes a non-symmetric uniaxial tensor, which can be written as $\overline{\varepsilon} = \varepsilon_t (\overline{I} - \hat{z}\hat{z}) + i\varepsilon_g (\hat{z} \times \overline{I}) + \varepsilon_a \hat{z}\hat{z}$, where $\varepsilon_g$ is the magnitude of the so-called gyration pseudovector, which vanishes if the bias is switched off (throughout this paper, we assume monochromatic fields with time-harmonic dependence $e^{-i\omega t}$). The frequency-dispersion model (magnetized Drude model) for all the elements of the tensor can be found, for example, in [19].

The propagation properties of surface and bulk modes are quite different on different planes orthogonal to the interface, depending on the angle of the plane with the static bias. SPPs are transverse-magnetic (TM) modes with respect to the propagation direction $\text{Re}[\mathbf{k}]$, with the electric field elliptically-polarized in a plane that includes $\text{Re}[\mathbf{k}]$ and the surface normal [20]. Since the Lorentz force acts on the charges *q* of the material as $\mathbf{F} \approx q\mathbf{E} + q\mathbf{v} \times \mathbf{B}_0$ (with **v** the carrier velocity), the SPPs that are mostly affected by the magnetization are those with charges moving in the plane orthogonal to the bias $\mathbf{B}_0$, namely, SPPs propagating in the *xy*-plane of Fig. 1. Thus, in this work we focus on TM wave propagation on this plane. We would like to stress, however, that while our analysis is in two dimensions, the considered plasmonic structure is still three-dimensional, and invariant along the *z*-axis. This scenario is very different compared to the case of a nonreciprocal two-dimensional plasma [5] (e.g., magnetized graphene at infrared frequencies [6]), in which the lateral confinement drastically changes the dispersion of the surface (edge) modes compared to the three-dimensional case considered here. We refer the reader to Ref. [5] for a complete analysis of magneto-plasmons in a magnetized two-dimensional electron gas.

How the magnetic bias affects the SPPs also strongly depends on the specific isotropic material above the interface and on whether the considered frequency $\omega$ is above or below the plasma frequency $\omega_p$.

All the cases of interest are illustrated in Fig. 1, with the typical dispersion diagram for TM bulk and surface modes schematically shown in Fig. 1(c,d). Two main classes of bound SPPs can be identified:

(I) When the material above the magnetized plasma is a transparent dielectric with permittivity $\varepsilon_d \geq \varepsilon_0$ (e.g., vacuum), bound surface waves are typically supported in the lower bulk-mode bandgap, as in the reciprocal case. However, nonreciprocity makes the SPP dispersion strongly asymmetrical, with the positive and negative branches converging to different flat asymptotes, which defines a frequency window where SPP propagation is allowed only in one direction [yellow-gray area in Fig. 1(c)]. While these surface magneto-plasmons have been studied for several decades [3], they have been recently shown to lose their strict unidirectional nature if nonlocal effects are included [14], as mentioned above.

(II) When the material above the magnetized plasma is, instead, an opaque isotropic medium with $\varepsilon_d < 0$ (e.g., a different unbiased metal, or a perfect electric conductor), the dispersion of the SPP modes changes dramatically, with two different frequency windows in which unidirectional surface waves are supported, as illustrated in Fig. 1(d). While the existence of these one-way surface waves has been known since at least the 1960s [1,2], their unidirectional nature has been recently interpreted in terms of certain topological properties of the bulk modes [4,7,15]. Indeed, it can be shown that the upper bulk-mode bandgap (opened due to time-reversal-symmetry breaking) is characterized by a non-zero topological invariant number, the so-called gap Chern number. Then, as in the case of quantum-Hall topological insulators in condensed-matter physics [21], topologically-protected unidirectional surface waves may emerge when an interface is formed with a material having a bulk-mode bandgap with different topological properties (the opaque isotropic material in our case) [4,7]. Conversely, the first type of SPPs discussed above, which exist on an interface with a transparent medium, are not topologically-protected, as the lower bulk-mode bandgap is topologically trivial [9].

The topological nature of this second type of SPPs seems to suggest that they may be more robust to the effect of nonlocality. Interestingly, it was shown in [4] that a high spatial-frequency cutoff (due to nonlocality) for the material response is *necessary* to justify, formally, the topological properties of continuum plasmonic media (a local material is not sufficiently well-behaved for infinite wavenumber, and the resulting band Chern number may not be an integer). However, the actual impact of nonlocality on SPPs of this type has not been investigated yet. In the following, we discuss how nonlocal effects influence the dispersion of both classes of SPPs described above, and we clarify whether truly unidirectional surface waves actually exist in these configurations.

## 3. IMPACT OF NONLOCALITY ON BULK MODES

The nonlocal response of a plasmonic material is due to the movement of free electrons during an optical cycle caused by convection and diffusion, which act to homogenize any inhomogeneity in the electron density [22]. This determines a spatial impulse response for the material that is not a delta function in space and, therefore, it becomes an explicit function of the wavevector $\mathbf{k}$ in spatial Fourier domain. Here, we treat nonlocal effects based on the hydrodynamic model of a free-electron gas, without diffusion effects, as done in [14] (we have verified that the inclusion of a small diffusion term does not qualitatively change our results). Using this approach, the problem of wave propagation in a nonlocal magnetized plasma can be solved by simultaneously solving Maxwell's equations, $\nabla \times \mathbf{H} = -i\omega\varepsilon_0\varepsilon_\infty \mathbf{E} + \mathbf{J}$, $\nabla \times \mathbf{E} = i\omega\mu_0 \mathbf{H}$, with induced free-electron current $\mathbf{J}$, and a microscopic hydrodynamic equation of motion for the free electrons that includes a pressure term determining convective currents (derived from the second moment of the Boltzmann transport equation). After some approximations and linearization, an independent equation for the induced current is obtained [14,22],

$$\beta^2 \nabla(\nabla \cdot \mathbf{J}) + \omega(\omega + i\gamma)\mathbf{J} = i\omega(\omega_p^2 \varepsilon_0 \varepsilon_\infty \mathbf{E} - \mathbf{J} \times \omega_c \hat{z}), \tag{1}$$

where $\beta$ is the nonlocal parameter, $\gamma$ is a phenomenological damping rate (absorption loss), $\varepsilon_\infty$ is the "background" relative permittivity representing the dielectric response due to bound charges, and $\omega_c$ is the plasma cyclotron frequency. By solving these equations simultaneously (details in Supplementary Material), the TM bulk-mode dispersion equation for a nonlocal magnetized plasma (in the lossless limit) can be found as

$$-k^4\beta^2\omega^2 + k^2\left[\Omega_b(\omega_p^2 + \beta^2 k_0^2 \varepsilon_\infty) + \Omega_0\right] - k_0^2 \varepsilon_\infty \Omega_0 = 0, \quad (2)$$

where $k_0 = \omega/c$, $\Omega_b = \omega^2 - \omega_p^2$ and $\Omega_0 = \Omega_b^2 - \omega^2\omega_c^2$. Interestingly, we note that the presence of the nonlocal parameter transforms the equation from quadratic to quartic with respect to $k$. This implies that, in a nonlocal plasma, two pairs of bulk modes (with equal and opposite $k$) are allowed to exist at any frequency $\omega$, compared to the single pair of modes in the local case with $\beta = 0$. This fact has important consequences, as we discuss in the following.

As a concrete example of nonlocal magnetized plasma, we consider n-doped indium antimonide (InSb), a material often considered as a model system in nonreciprocal plasmonics [11,14,23]. The parameters of a typical sample of this material are the following: effective mass $m^*/m_0 = 0.0142$, electron density $N_e = 1.1 \times 10^{22}$ m$^{-3}$, which yields a plasma frequency $\omega_p/2\pi = 2$ THz, and background permittivity $\varepsilon_\infty = 15.6$ [11,24]. Thanks to the small effective mass of InSb, a significant cyclotron frequency with respect to the plasma frequency, $\omega_c/\omega_p = B_0\sqrt{\varepsilon_0\varepsilon_\infty/m^*N_e}$, can be obtained even with moderate bias intensity. However, since the nonlocal parameter is proportional to the Fermi velocity, $\beta^2 = 3v_F^2/5$, which is inversely proportional to the effective mass [25], InSb also exhibits a moderately large nonlocal response, as recognized in [14]. Conversely, gas plasmas would have much weaker nonlocality due to their effective mass $m^* \approx m_0$. Figure 2(a) shows the TM bulk-mode dispersion diagram, for wave propagation normal to the bias, in magnetized InSb with $\omega_c = 0.2\omega_p$, for different values of the nonlocal parameter. As clearly seen in the figure, while in the local case the bulk-mode dispersion exhibits a full bandgap for any value of $k$, owing to the flat dispersion of the lower bulk mode, if the nonlocal parameter is non-zero the bandgap "exists" only for a limited range of wavenumbers since the lower bulk mode now tends to infinite $\omega$ for infinite $k$, as predicted by Eq. (2). We also note that the bandgap does not "close" in the conventional sense, since the two bulk modes never cross as they tend to infinite frequency.

Moreover, we see that, even for the large nonlocal parameter of InSb, $\beta = 1.07 \times 10^6$ m/s, the nonlocal dispersion drastically departs from the local one only for large values of $k$, in the order of a hundred times the free-space wavenumber, at the limit of validity of the hydrodynamic nonlocal model. Indeed, the Boltzmann transport equation, from which the hydrodynamic model is derived, is a semi-classical equation where the distribution function of the fermions is treated as classical (the Heisenberg uncertainty principle is not considered, and position and momentum are assumed to be known with the same accuracy). This means that the hydrodynamic model is invalid if the wavelength of the relevant electromagnetic modes of the system is comparable to the Fermi wavelength (i.e., the wavelength of a fermion at the Fermi energy) [25]. For the considered electron density of n-type InSb, the Fermi wavelength is $\lambda_F = 2\pi/(3\pi^2 N_e)^{1/3} = 9.1 \times 10^{-8}$ m. Thus, we can define an approximate validity bound for the hydrodynamic model, for the considered material, by requiring that the wavelength of the bulk/surface modes is sufficiently larger than the Fermi wavelength. We choose $\lambda > 10\lambda_F$, corresponding to $|k| < 0.1 \times 2\pi/\lambda_F \approx 10^6$ m$^{-1}$, which leads to a valid wavenumber range $|k| < 150 k_p$, where $k_p$ is the free-space wavenumber at the InSb plasma frequency. Any analysis of InSb optical nonlocality based on the hydrodynamic approach is therefore valid within this range, which is, however, sufficiently wide to allow using this model to make relevant predictions for the optical response of the considered material (indeed, modes with even large wavenumber would be very weakly excited in any practical configuration, and can typically be neglected).

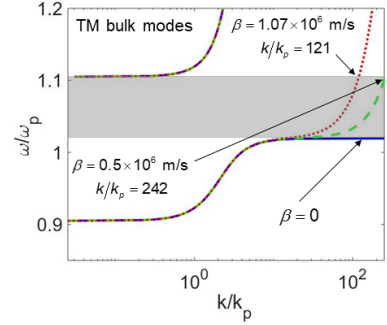

Fig. 2. Dispersion diagram of the TM bulk modes for a homogenous sample of magnetized InSb, with $\omega_p/2\pi = 2$ THz, $|\omega_c|/\omega_p = 0.2$, and $\gamma = 0$, in the plane orthogonal to the bias. Different curves are for increasing values of the nonlocal parameter, as indicated in the figure (the second number in the labels is the value of $k$ at which the lower bulk mode reaches the frequency of the upper bulk mode). $k_p$ is the free-space wavenumber at the plasma frequency.

## 4. IMPACT OF NONLOCALITY ON SURFACE MODES

We now focus on how nonlocal effects influence the dispersion of SPPs of both types introduced in Section 2. The surface-mode dispersion equation can be found analytically, as detailed in Supplementary Material, by imposing the continuity of the tangential fields at the interface, with additional boundary conditions due to presence of additional bulk modes [22,26]. The specific form of the additional boundary conditions depends on the specific material above the interface, as discussed below.

We first briefly consider the case of conventional surface magneto-plasmons [type-I SPPs in Fig. 1(a,c)], to confirm some results recently reported in the literature [14] and for comparison with the other cases of interest. Since the material above the magnetized plasma is a dielectric, an additional boundary condition can be defined by requiring the free-electron current normal to the surface to vanish at the plasma-dielectric interface, i.e., $\hat{y} \cdot \mathbf{J} = 0$. The resulting SPP dispersion, for both the local and nonlocal cases, is reported in Fig. 3(a), within the validity range of the hydrodynamic model. The flat asymptotic behavior of the local dispersion curves is nonphysical since it violates thermodynamics, as demonstrated in [14] (an infinite number of photonic states within a finite frequency range implies infinite energy density at a finite temperature). Indeed, this nonphysical behavior immediately disappears if non-zero nonlocality is considered, as seen in Fig. 3(a), with the dispersion curves now monotonically growing for large $k$. This fact also directly implies the absence of truly unidirectional type-I SPPs: at every frequency, SPP modes are allowed to propagate in both directions, with different wavenumbers. In the nonlocal case, therefore, a point source, exciting a wide-enough range of spatial frequencies, is able to launch surface waves in both directions, as confirmed by our numerically-calculated field distributions in Fig. 3(b-f) [Eq. (1) was implemented as a weak-form equation in a commercial finite-element software [27]). A backward highly-oscillating mode (large $k$) is clearly visible in the nonlocal low-loss case [Fig. 3(c,e); right-going SPP], confirming the conclusions of [14]. Moreover, we note that the strength of the backward mode strongly depends on the level of absorption losses: as seen in Fig. 3(d,f), a moderate level of loss is sufficient to suppress the right-going SPP, restoring the unidirectionality, while the left-going SPP is only moderately attenuated.

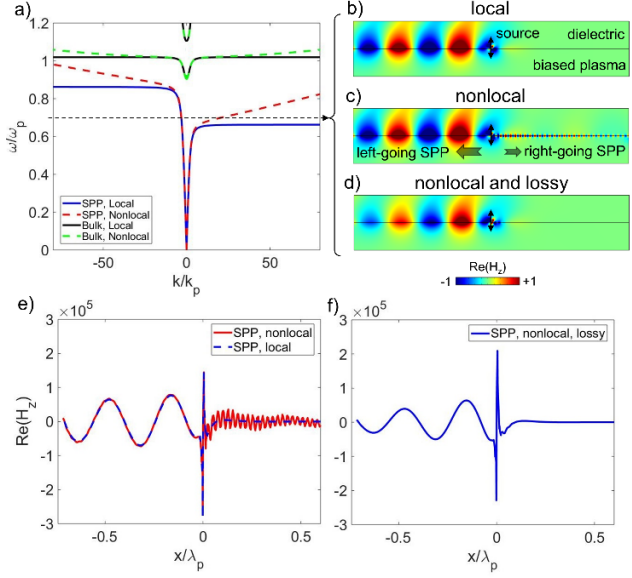

Fig. 3. (a) Dispersion diagram of type-I SPPs at the interface between magnetized InSb, with $|\omega_c|/\omega_p = 0.2$, and a dielectric material (silicon, with $\varepsilon_d = 11.68$), for both local ($\beta = 0$) and nonlocal ($\beta = 1.07 \times 10^6$ m/s) plasma models. (b,c,d) Real part of the magnetic-field distributions (time-snapshots) for type-I SPPs launched by a dipolar source (black arrow) for different cases at $\omega/\omega_p = 0.7$: (b) local and low-loss plasma, with $\gamma = 0.002\omega_p$; (c) nonlocal and low-loss plasma [$\gamma$ as in (b)]; (d) nonlocal plasma with moderately-high loss, $\gamma = 0.05\omega_p$. (e,f) one-dimensional field profiles, corresponding to (b,c,d), at a distance of $\approx \lambda_p/100$ below the interface.

The situation is drastically different for type-II SPPs, which are supported by an interface between a magnetized plasma and an opaque isotropic material, at frequencies above $\omega_p$ [Fig. 1(b,d); upper SPP branch]. The opaque material may be another unbiased plasmonic medium, with plasma frequency $\omega_p^m > \omega_p$, such that its permittivity is negative across the bulk-mode bandgap of the magnetized plasma. We consider a nonlocal hydrodynamic model for the unbiased plasmonic material as well, which yields an additional pair of bulk modes, as in Eq. (2), and requires including two additional boundary conditions to solve for the surface modes. Clearly, the condition $\hat{y} \cdot \mathbf{J} = 0$ cannot be used at a plasma-plasma interface; instead, following Ref. [26], we enforce the continuity of hydrodynamic pressure and normal free-electron velocity at the interface, which are physically-meaningful conditions as extensively discussed in [26]. Further details are provided in Supplementary Material. Figure 4(a) shows the resulting SPP dispersion for both local and nonlocal cases, similar to Fig. 3(a). First, we note that this surface mode exhibits a lower-frequency cutoff at $k = 0$ and $\omega = \sqrt{\omega_c^2 + \omega_p^2}$ (the modal wavenumber is purely imaginary below this frequency, in the lossless case). Instead, at these frequencies, this SPP does not show any nonphysical flat asymptotic dispersion even in the local scenario, and it exists for low values of wavenumber, where the effect of nonlocality, proportional to $\beta^2 k^2$ according to Eq. (1), is expected to be small. Indeed, our exact calculations in Fig. 4(a) show that the dispersion of type-II SPPs is completely unaffected by nonlocality at frequencies around the bulk-mode bandgap, and no backward mode emerges within the validity range of the hydrodynamic model (more details in the following). The unidirectionality of these surface waves is further confirmed by the numerically-calculated magnetic-field distribution of SPPs launched by an ideal dipolar source, reported in Fig. 4(b-e), which shows no sign of a backward-propagating mode, in drastic contrast with the behavior of type-I SPPs in Fig. 3. This is a rather remarkable result, as it represents a clear example of unidirectional SPP, existing on a nonreciprocal plasmonic platform, that is very robust to the effect of spatial dispersion.

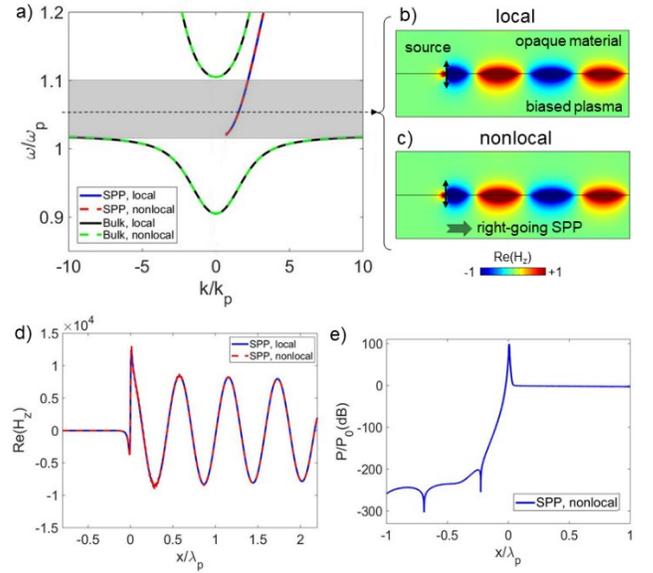

Fig. 4. Similar to Fig. 3, but for type-II SPPs at the interface between a magnetized InSb and an opaque isotropic material. The parameters of the biased plasma are the same as in Fig. 3. The opaque material is an unbiased plasmonic medium with $\omega_p^m = 2\omega_p$, $\varepsilon_\infty^m = \varepsilon_\infty$, and nonlocal parameter $\beta = \beta_m$. (a) Dispersion diagram. (b,c,d) Magnetic field-distributions (time-snapshots) at $\omega/\omega_p = 1.05$, for local and nonlocal cases. (e) Corresponding time-averaged power-flow distribution (absolute value of the real part of the complex Poynting vector along $x$), for the nonlocal case, demonstrating negligible power flow toward the negative $x$-axis.

While Fig. 4(a) highlights the most relevant region of the dispersion diagram for type-II unidirectional SPPs, in order to get a complete picture of all the SPP modes supported in a configuration with opaque isotropic cover [Fig. 1(b)], we also investigated the surface-mode dispersion over a much broader frequency-wavenumber range. Fig. 5 shows dispersion diagrams, for the local and nonlocal cases considered in Fig. 4, obtained by plotting the inverse determinant of the boundary-condition matrix of the system. The poles of this quantity (hence, the bright branches in Fig. 5) correspond to *all* the possible surface modes of the structure. Some relevant observations should be made about these dispersion diagrams. First, we note the presence of a second SPP mode that exists at very low frequencies ($\omega < \omega_c$), as sketched in Fig. 1(d). In the local case [Fig. 5(a,c)], this low-frequency SPP mode appears to be unidirectional; however, its flat asymptotic dispersion suggests that, just like type-I SPPs, it may be rather strongly affected by spatial dispersion. Indeed, as shown in Fig. 5(b,f), if nonlocal effects are introduced (same parameters as in Fig. 4), the mode loses its nonphysical flat dispersion and, moreover, it becomes bi-directional at all frequencies due to the emergence of another SPP branch with positive wavenumber [this additional SPP branch becomes a flat dispersion line at $\omega = 0$ in the local case]. In the presence of nonlocal effects, the dispersion of these low-frequency SPPs tends to infinite $\omega$ for infinite $k$, and, therefore, it reaches the frequencies of the type-II SPPs described above ($\omega > \omega_p$). However, it is evident from Fig. 5(b) that, at such frequencies, the wavenumber of these additional SPP modes is extremely large for any realistic value of nonlocality, significantly beyond the limits of validity of the hydrodynamic model, which makes them essentially impossible to excite by any source or discontinuity (their wave impedance, $\propto k_x$, is very large), as also confirmed by our full-wave simulations in Fig. 4.

The plots in Fig. 5 also gives us a better picture of type-II SPPs over a wider frequency range around the bulk-mode bandgap, and offer additional insight into their behavior. As clearly seen, in both the local [Fig. 5(d)] and nonlocal [Fig. 5(e)] case, the SPP mode is unidirectional within the bandgap, as discussed above, whereas a backward-propagating SPP mode appears right above the bandgap. This backward type-II SPP is typically not shown in the literature (e.g., [7]), and we don't show it in Fig. 1(d) and 4(a), because, at frequencies close to the bandgap, it is almost perfectly overlapped to the dispersion curve of the upper bulk mode, and it goes under cutoff within the gap. It is indeed the difference in lower-frequency cutoffs between backward- and forward-propagating SPPs of this type, due to the magnetic bias,
that opens a unidirectional frequency range for type-II SPPs. The cutoff-frequency difference can be found analytically, and written as

$$\Delta = \omega_c/2 + \sqrt{\omega_c^2/4 + \omega_p^2} - \sqrt{\omega_c^2 + \omega_p^2} , \qquad (3)$$

which is equal to the width of the bulk-mode bandgap, and vanishes if $\omega_c = 0$, as expected. $\Delta$ is also completely independent of the nonlocal parameters since the modes go under cutoff at $k = 0$. This mechanism for unidirectionality, based on a biased-induced *asymmetry in the lower-frequency cutoffs*, is much more robust to nonlocality compared to the mechanism behind unidirectional type-I SPPs, i.e., an *asymmetry in the flat asymptotic dispersion* for $k \to \infty$, which is unphysical.

The effect of nonlocality on type-II SPPs is only visible at higher frequencies, as the dispersive negative permittivity of the isotropic cover becomes smaller. In the local case [Fig. 5(d)], the SPP modes exhibit asymmetric flat bands for large $k$, similar to the behavior of type-I SPPs near the surface-plasmon resonance. Again, this unphysical behavior is removed when nonlocal effects are included, as seen in Fig. 5(e); however, we also note that the dispersion lines in the nonlocal case become blurred, as the SPP pole rapidly moves to complex values of frequency/wavenumber. Since we assumed no absorption losses in Fig. 5, this behavior indicates that the SPP mode decays radiatively as a leaky mode [28]. Indeed, due to nonlocality, the magnetized plasma supports a second large-$k$ bulk mode at these frequencies (Fig. 2), which provides a continuum of outgoing waves for the SPP to couple to at different angles (based on transverse momentum matching), just like for leaky waves in free space. This nonlocality-induced radiation leakage into the bulk was already recognized in [26] at an interface between two metals, but, to the best of our knowledge, this is the first time this behavior is predicted for a nonreciprocal plasmonic configuration.

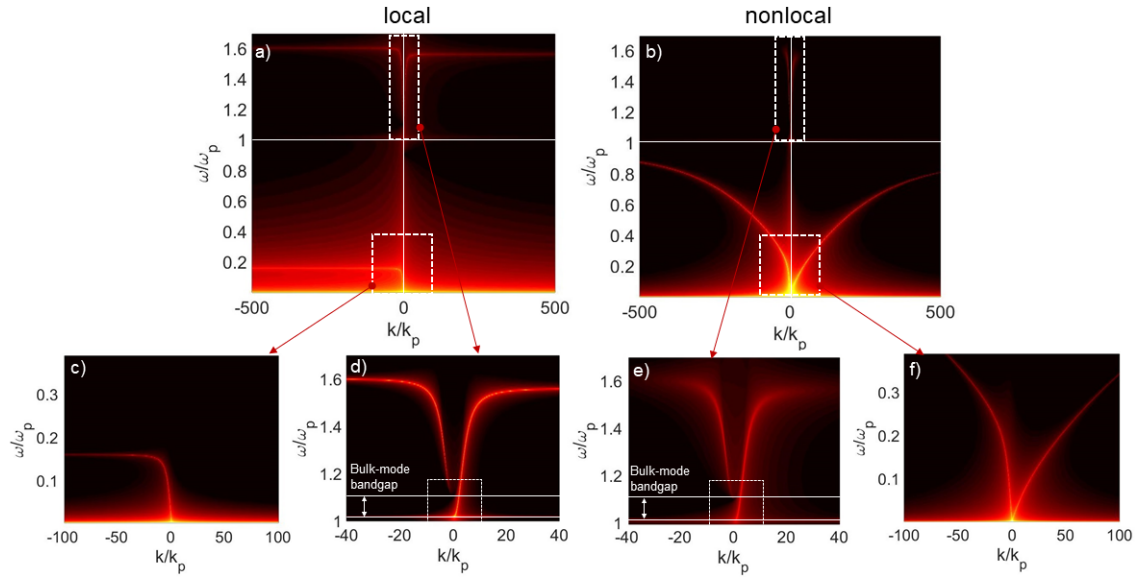

Fig. 5. Dispersion diagram of type-II SPPs, for the local and nonlocal configurations considered in Fig. 4, over a much broader range of frequencies and wavenumbers. To capture all the surface modes, the dispersion diagrams are plotted as density plots of the inverse determinant of the boundary-condition matrix. The bright lines correspond to the SPP poles. (a) Local case, with zoomed-in views (dashed white boxes) of the ultra-low-frequency SPP (c), and of the SPP branches above the plasma frequency (d). (b,e,f) Corresponding plots for the nonlocal case. The white solid lines in (d) and (e) indicate the bulk-mode bandgap. The dashed white boxes in (d) and (e) correspond to the dispersion diagram in Fig. 4(a).

## 5. TERMINATED ONE-WAY CHANNELS

An interesting configuration that reveals the counter-intuitive behavior of unidirectional waves is the case of a *terminated* one-way channel, which can be realized here by introducing an impenetrable wall orthogonal to the magnetized plasma interface, as considered, e.g., in [2,11]. For type-I SPPs the termination should be a perfect electric conductor (PEC) since, for $\omega < \omega_p$, no surface wave is supported by an interface between a magnetized plasma and a PEC wall. Hence, a SPP wave incident on the termination cannot travel along the termination itself; however, it can still "escape" via different channels: absorption loss in the materials, radiation loss into the transparent cover, as well as through a backward-propagating SPP mode, which exists at any frequency if nonlocal effects are correctly accounted for, as discussed above. The excitation of a backward wave is clearly seen in our full-wave simulations in Fig. 6(a,b,c), consistent with the results in [14].

Again, the situation is drastically different for type-II SPPs, within the bulk-mode bandgap. In this case, the correct termination is a perfect magnetic conductor (PMC), which does not support any surface wave at an interface with a magnetized plasma [2] (whereas a PEC-plasma interface would support surface-wave propagation at these frequencies). In drastic contrast with the previous case, in this configuration the energy carried by an incident type-II SPP cannot escape through a backward-propagating mode, as discussed in the previous section. In addition, no radiation loss in the cover is possible since the material is opaque, while scattering/leakage into the bulk modes of the nonreciprocal plasma is minimal at these frequencies and becomes negligible if nonlocal effects are weak (the SPP pole is essentially on the real axis). As seen in the field distributions in Fig. 6(d,e,f), the SPP wave impinges on the interface with no reflection [Fig. 6(d)], accompanied by a very rapid increase in field intensity at the corner [Fig. 6(e,f)], while its energy is dissipated via absorption. Using transmission-line terminology, this type of one-way channel provides *automatic impedance-matching for any load*, including for a load with vanishing resistive part (vanishing absorption loss). We also would like to note that the ideal lossless case does not lead to any thermodynamic paradox, as demonstrated by Ishimaru in Ref. [2] within the framework of macroscopic electrodynamics, without the need for any microscopic description of the involved materials. He showed that a terminated one-way channel, in the ideal lossless scenario, is an improperly-posed boundary-value problem since the fields are not integrable at the termination and the solution is discontinuous. In fact, the integral form of the Poynting theorem is meaningful and yields real non-zero power dissipation in the limit of vanishing losses, but not in the ideal lossless case [2]. This is also nicely discussed in a recent paper by Mann, et al. [13].

Moreover, we note that the presence of very intense fields near the termination (arbitrarily large in the case of vanishing losses), accompanied by a continuous shrinking of the SPP wavelength at the corner, is similar to the behavior of wedge-like reciprocal plasmonic structures [20], but this behavior is facilitated here by the unidirectional nature of the involved SPP mode, which eliminates the need for adiabatic tapering to ensure impedance matching. The origin and nature of unidirectionality-enhanced "field hotspots" in nonreciprocal plasmonic structures is also nicely discussed in [16] for local materials. Most importantly, our results in this Article show that the presence of nonlocal effects does not prevent the formation of such field hot spots for type-II SPPs, as clearly seen in Fig. 6(f). Conversely, nonlocality does weaken the field intensity of type-I SPPs at the termination significantly [Fig. 6(c)] due to the emergence of a moderately-strong backward-propagating mode that provides an additional "escape" channel. We speculate that the dramatic field-intensity enhancement that can be achieved in terminated one-way channels, based on robust type-II SPPs, may be useful to locally enhance light-matter interactions at the subwavelength scale by orders of magnitude.

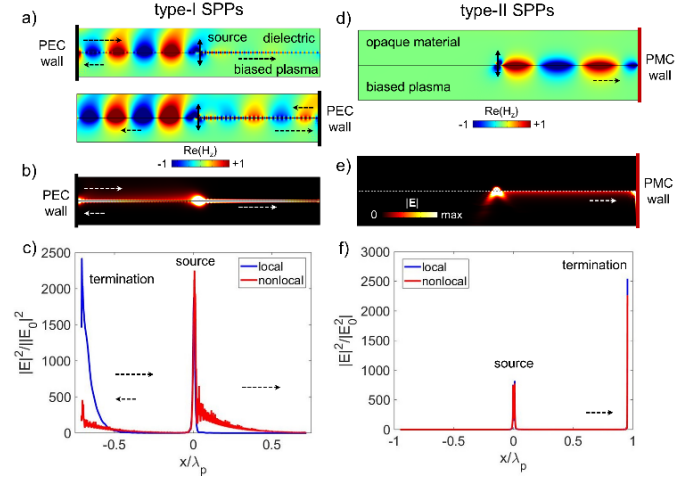

Fig. 6. Surface-wave propagation on a *terminated* interface between a magnetized plasma and an isotropic material. Left column: type-I SPPs, with PEC termination. Same parameters as in Fig. 3 for the nonlocal low-loss case. Right column: type-II SPPs, with PMC termination. Same parameters as in Fig. 4 for the nonlocal low-loss case. (a,d) Real part of the magnetic-field distributions (time-snapshots) for SPPs launched by a dipolar source. (b,e) Corresponding electric-field intensity distributions. (c,f) Field-intensity profiles, corresponding to (b,e), at a distance $\approx \lambda_p/100$ below the surface, normalized by the field intensity of a SPP on a lossless homogenous interface. Field profiles for the local case are also shown in (c,f) for comparison. Longer/shorter arrows indicate the propagation direction of SPPs with larger/smaller wavenumber. Time-harmonic animations for (a,d) are available as Supplementary Material.

## 6. CONCLUSION

In summary, we have provided a comprehensive analysis of surface-plasmon-polariton modes existing on the surface of a three-dimensional nonreciprocal plasmonic platform in different scenarios. We have shown that an important class of unidirectional surface plasmon-polaritons, existing at an interface between a magnetized plasma and an opaque isotropic material for $\omega > \omega_p$, are robust to nonlocal effects since their unidirectionality does not depend on an unphysical dispersion at large wavenumbers. Instead, their unidirectional frequency window is based on a bias-induced asymmetry between lower-frequency cutoffs for backward- and forward-propagating modes at $k = 0$. Our results also confirm that other classes of SPPs, including conventional surface magneto-plasmons, clearly become bi-directional if nonlocal effects are considered.

In general, our findings clarify that the impact of spatial dispersion in nonreciprocal plasmonics strongly depends on the considered configuration and frequency regime, and that nonlocal effects do not rule out unidirectionality in all cases. We expect that our work will stimulate intensive research to observe and exploit robust unidirectional surface plasmon-polaritons in nonreciprocal plasmonic platforms.

See Supplement 1 for supporting content.